\documentclass[aip,cha, reprint,showpacs,showkeys]{revtex4-1}
\usepackage{amssymb, dcolumn,  graphicx, latexsym, amsfonts, bm, float}

\begin{document}

\title{Drift–Alfven instabilities of a finite beta plasma sheared flow along a magnetic field with inhomogeneous ion temperature}
\author{V. V. Mikhailenko}\email[E-mail: ]{vladimir@pusan.ac.kr}
\affiliation{Plasma Research Center, Pusan National University,  Busan 46241, South Korea.}
\affiliation{BK21 Plus Information Technology, Pusan National University,  Busan 46241, South Korea.}
\author{V. S. Mikhailenko}\email[E-mail: ]{vsmikhailenko@pusan.ac.kr}
\affiliation{Plasma Research Center, Pusan National University,  Busan 46241, South Korea.}
\author{Hae June Lee}\email[E-mail: ]{haejune@pusan.ac.kr}
\affiliation{Department of Electrical Engineering, Pusan National University, Busan 46241, South Korea.}

\date{\today}

\begin{abstract}
The drift--Alfven instabilities in the magnetic field aligned (parallel) sheared flow of a finite beta $(1 > \beta > m_{e}/m_{i})$ plasma with 
comparable inhomogeneous ion temperature and homogeneous electron temperature are examined. The development of  instabilities are 
quantitatively discussed on the basis of numerical solution of a set of equations for the electrostatic and electromagnetic potentials. 
It is found that the accounting for the electromagnetic ion kinetic response, which has been ignored 
usually in existing discussions of the drift-Alfven instabilities of a steady plasma, reveals new drift-Alfven instability driven by the coupled action of 
the ion temperature gradient, the flow velocity shear, and the ion Landau damping. The excited unstable waves have the phase
velocities along the magnetic field comparable with the ion thermal velocity, and the growth rate comparable with the frequency. 
\end{abstract}

\maketitle

\section{Introduction}\label{sec1}
The understanding of instabilities driven by a shear in the velocity of a plasma flows is of considerable interest to research  in magnetic 
fusion plasmas, where inhomogeneous flows and currents along and across magnetic field are ubiquitous. It is well known, that the 
magnetic field aligned plasma shear flows, which are observed in the edge layers of tokamak plasmas\cite{Asakura, LaBombard, Fedorczak, Pedrosa, Wang}  
are the additional sources of free energy for the electrostatic and electromagnetic instabilities development. The majority of analysis of these 
instabilities have been restricted to the case of cold ions with ion temperature $T_{i}$ much less than the electron temperature $T_{e}$. In this case
the analysis may be simplified by considering the ions in the fluid limit which is valid only for the perturbations having the phase velocity along
the magnetic field that greatly exceeds the thermal velocity of the ions $v_{Ti}$. For the tokamak plasma, the ionosphere and the solar wind  plasmas the 
case of warm ions having the temperature comparable with or even exceeding the electron temperature is more relevant.
 
In parallel shear flows with hot ions, $T_{i}\gtrsim T_{e}$, the ion kinetic effects play the decisive role in the development the instabilities of the 
parallel shear flows. In particular, such plasma is prone to the excitation of the shear-flow driving instabilities \cite{Mikhailenko-2014, 
Mikhailenko-2016}. It was obtained \cite{Mikhailenko-2016}, that in plasma with inhomogeneous density the ion kinetic drift--Alfven (DA) instability develops 
from the coupled action of the parallel--flow shear and ion Landau damping. The growth rate of this instability is of the order of the frequency, the phase 
velocity of perturbations along the magnetic field is comparable with ion thermal velocity. The electromagnetic response of the ions for this instability 
is comparable with the electromagnetic response of the electrons. 

It was found in Ref.\cite{Mikhailenko-2014}, that the electrostatic instability exists by virtue of the coupled action of parallel velocity 
shear, ion Landau damping and a ion temperature gradient, which reinforce each other in the development of the ion kinetic instability. In this paper we 
undertake the investigation of the electromagnetic counterpart of that instability: the DA instability of the parallel shear flow with 
inhomogeneous ion temperature.  In Section \ref{sec2}, we present the basic linear equations that govern the stability properties of the parallel shear 
flows with inhomogeneous flow density and inhomogeneous ion temperature. In Section \ref{sec3}, we present the numerical solution and discussion of the 
detected electromagnetic  DA instabilities with special emphasis on the thermal and  electromagnetic effects of ions on the stability properties 
of the parallel shear flows. Conclusions are given in Section \ref{sec4}.

\section{Basic equations}\label{sec2}

The tokamak plasma is, as a rule, a low $\beta$ plasma, with $1 \gg \beta \gg m_{e}/m_{i}$. It is well known, that such a plasma is unstable 
against the development of the electromagnetic DA instabilities\cite{Mikhailovskii}. These instabilities are governed by the 
Vlasov equations for electrons and ions and the Poisson equation and Ampere's law for the electrostatic 
potential $\Phi$ and along magnetic field component $A_{z}$ of the electromagnetic potential. For the inhomogeneous, magnetic-field-aligned,
single--ion--species, collisionless plasma flow with velocity ${\bf{V}}_{0}(X_{\alpha})\parallel B_{0} {\bf{e}}_z $, 
the Vlasov equation for the perturbation $f_{\alpha}=F_{\alpha}-F_{0\alpha}$ of the distribution function $F_{\alpha}$ with 
equilibrium function $F_{0\alpha}$ in  guiding center coordinates in slab geometry,
$X_{\alpha}=x+\frac{v_{\bot}}{\omega_{c\alpha}}\sin\phi$, $Y_{\alpha}=y-\frac{v_{\bot}}{\omega_{c\alpha}}\cos\phi$,  
where $x$ and $y$ are coordinates of the particle position, $\omega_{c\alpha}$ is the cyclotron frequency, has a form
\begin{eqnarray}
& \displaystyle 
\frac{\partial f_{\alpha}}{\partial t}-\omega_{c\alpha}
\frac{\partial f_{\alpha}}{\partial\phi} +v_{z}\frac{\partial
f_{\alpha}}{\partial z} 
\nonumber
\\ 
& \displaystyle 
=\frac{e}{m_{\alpha}}\left[ \frac{1}{\omega_{c\alpha}}\left( \frac{\partial\Phi}{\partial Y}
-\frac{v_{z}}{c}\frac{\partial A_{\parallel}}{\partial Y}\right) 
\frac{\partial F_{0\alpha}}{\partial X}\right. 
\nonumber
\\ 
& \displaystyle
-\frac{\omega_{c\alpha}}
{v_{\bot}}\left( \frac{\partial\Phi}{\partial \phi} 
-\frac{v_{z}}{c}\frac{\partial A_{\parallel}}{\partial \phi}\right) \frac{\partial F
_{0\alpha}}{\partial v_{\bot}} 
\nonumber
\\ 
& \displaystyle
\left. 
+\left( \frac{\partial\Phi}{\partial z} +\frac{1}{c}\frac{\partial A_{\parallel}}{\partial t}
-\frac{\omega_{c\alpha}}{c}\frac{\partial A_{\parallel}}{\partial \phi}\right) 
\frac{\partial F_{0\alpha}}{\partial v_{z}}\right],
\label{1}
\end{eqnarray}

The perturbed electrostatic potential $\Phi$ is determined by the Poisson equation
\begin{eqnarray}
&\displaystyle 
\Delta \Phi\left(\mathbf{r},t\right)=
-4\pi\sum_{\alpha=i,e} e_{\alpha}\int f_{\alpha}\left(\mathbf{v},
\mathbf{r}, t \right)d\mathbf {v}_{\alpha}. \label{2}
\end{eqnarray}
and perturbed electromagnetic potential $A_{\parallel}$ is determined by the Ampere's law
\begin{eqnarray}
&\displaystyle 
\Delta A_{\parallel}\left(\mathbf{r},t\right)=
-\frac{4\pi}{c}\sum_{\alpha=i,e} e_{\alpha}\int v_{z} f_{\alpha}\left(\mathbf{v},
\mathbf{r}, t \right)d\mathbf {v}_{\alpha}. \label{3}
\end{eqnarray}
\begin{figure}[!htbp]
\includegraphics[width=0.4\textwidth]{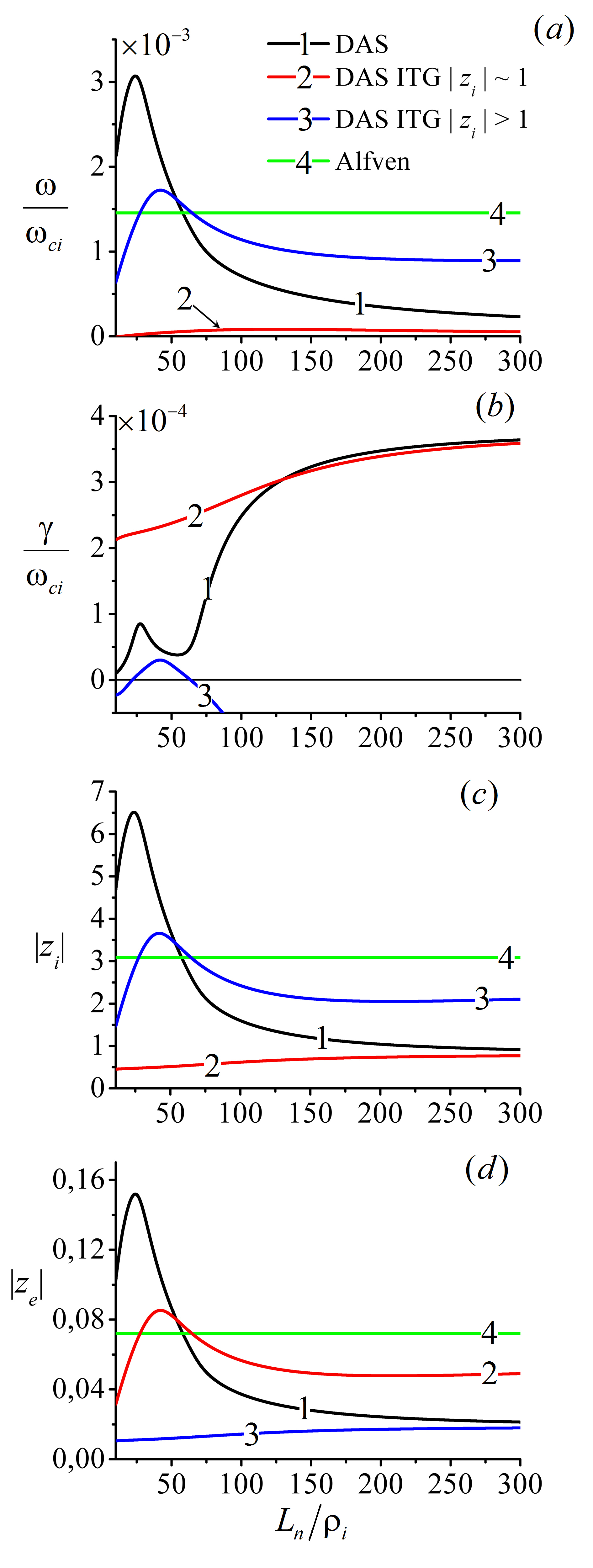}
\caption{\label{fig1} The normalized frequency $\omega/\omega_{ci}$ (panel (a)), the normalized growth rate $
\gamma/\omega_{ci}$ (panel (b)), $\left|z_{i} \right|$ (panel (c)) and $\left|z_{e} \right|$ (panel (d))
versus $L_{n}/\rho_{i}$ for $\eta_{i} = 3$, $k_{y}\rho_{i} = 0.1$, $T_{i}/T_{e} = 
1$, $\left( V'_{0}/\omega_{ci}\right) ^{-1} =70$, $\left( k_{z}\rho_{i}\right) ^{-1} = 3000$.}
\end{figure}
\begin{figure}[!htbp]
\includegraphics[width=0.4\textwidth]{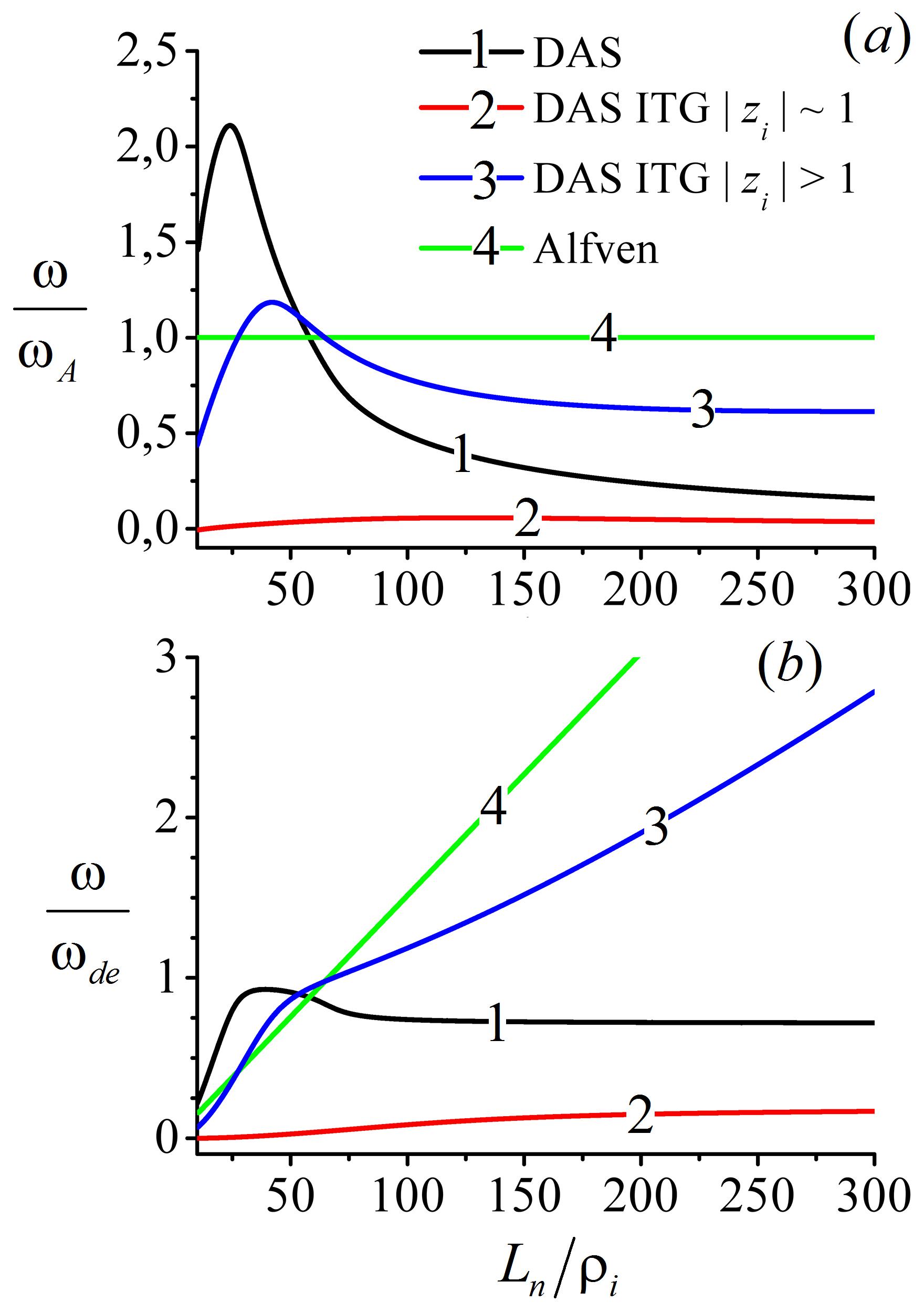}
\caption{\label{fig2} The normalized frequencies $\omega/\omega_{A}$ (panel (a)) and $\omega/\omega_{de}$ (panel (b))
versus $L_{n}/\rho_{i}$ for $\eta_{i} = 3$, $k_{y}\rho_{i} = 0.1$, $T_{i}/T_{e} = 
1$, $\left( V'_{0}/\omega_{ci}\right) ^{-1} =70$, $\left( k_{z}\rho_{i}\right) ^{-1} = 3000$.}
\end{figure}

In what follows, $F_{0\alpha}$ is considered as the shifted Maxwellian 
distribution function for electrons and ions ($\alpha=i,e$)
\begin{eqnarray}
& \displaystyle 
F_{0\alpha} = \frac{n_{0\alpha}\left(X_{\alpha}
\right)}{\left({2\pi v_{T\alpha}^2}\right)^{3 /2}}\exp\left[-
\frac{v_{\perp}^{2}}{2v_{T\alpha}^2} - \frac{\left(v_{z} -
V_{0}(X_{\alpha})\right)^2}{2v_{T\alpha}^{2}} \right],\label{4}
\end{eqnarray}
assuming the inhomogeneity direction of the density and temperature of the
sheared-flow species is along coordinate $X_{\alpha}$, $v_{T\alpha}=\left(T_{\alpha}
\left(X_{\alpha}\right)/m_{\alpha} \right)^{1/2}$ is the thermal velocity. The flow velocity 
of ions $\mathbf{V}_{0}$ is assumed to be equal to that of the electrons.  We consider here the 
idealized inhomogeneous-flow case of homogeneous parallel-velocity shear, i.e. 
$V_{0}(X_{\alpha})=V_{00}+V'_{0}X_{\alpha}$, where 
$V_{00}$ is the spatially homogeneous part of the flow velocity, and $V'_{0}=const$. In 
order to simplify the problem, a velocity $\mathbf{v}$ usually transforms from the 
laboratory to a convecting frame of reference, $\mathbf{v}= \hat{\mathbf{v}}+V_{0}
(X_{\alpha})\mathbf{e}_{z}$, where any spatially homogeneous part of flow velocity is 
eliminated from the problem by a simple Galilean transformation. The solution of the system (\ref{1})-(\ref{3}) in the 
convective set of reference is of the modal form during a long time until $V'_{0}t\lesssim 
k_{x}/k_{z}$\cite{Mikhailenko-2014}. Until that time, the  solution to Fourier transformed system (\ref{1})--(\ref{3}) in the local approximation, 
for which $k_xL_{n} \gg 1$, 
where $L_{n} = \left[ d\ln n_{0}\left(X \right)/dx \right]^{-1}$, is given by the equations
\begin{eqnarray}
& \displaystyle
\Phi\left(\mathbf{k},\omega \right) \left[k^{2}\lambda^{2}_{Di}
+\sum\limits_{n=-\infty}^{\infty}A_{ni}
\Big[1+i\sqrt{\pi}W\left( z_{ni}\right)\Big.
\right.
\nonumber
\\ & 
\displaystyle 
\times
 \left. \left(z_{0i}-\chi_{i} \left(1-\frac{1}{2} \eta_{i}\right) \right)  \right]
\nonumber
\\ 
& \displaystyle 
-S_{i}\sum\limits_{n=-\infty}^{\infty}A_{ni}
\Big(1+i\sqrt{\pi}z_{ni} W\left( z_{ni}\right)   \Big)
\nonumber
\\ 
& \displaystyle 
-\sum\limits_{n=-\infty}^{\infty}\chi_{i}\eta_{i}z_{ni}A_{ni}
\Big(1+i\sqrt{\pi}z_{ni} W\left( z_{ni}\right)   \Big)
\nonumber
\\ & \displaystyle 
+i\sqrt{\pi}\sum\limits_{n=-\infty}^{\infty}\chi_{i}\eta_{i}
 W\left( z_{ni}\right) k^{2}_{\perp}\rho_{i}^{2}\left(A_{ni}+\hat{A}_{ni}\right)  
\nonumber
\\ 
& \displaystyle
\left.
+\tau\sum\limits_{n=-\infty}^{\infty}A_{ne}
\Big(1+i\sqrt{\pi}\left(z_{0e}-\chi_{e} \right)W\left( z_{ne}\right)   \Big)
\right] 
\nonumber
\\ 
& \displaystyle
-A_{\parallel}\left(\mathbf{k},\omega \right)\left\{
\sum\limits_{n=-\infty}^{\infty}A_{ni}\left(z_{0i}-
\chi_{i}\left(1-\frac{1}{2} \eta_{i}\right) \right)\right. 
\nonumber
\\ & \displaystyle \times
\Big(1+i\sqrt{\pi}z_{ni} W\left( z_{ni}\right)   \Big)
\nonumber
\\ & \displaystyle
-S_{i}\sum\limits_{n=-\infty}^{\infty}A_{ni}
z_{ni}\Big(1+i\sqrt{\pi}z_{ni} W\left( z_{ni}\right)   \Big)
\label{5} 
\\ 
& \displaystyle 
-\sum\limits_{n=-\infty}^{\infty}\chi_{i}\eta_{i}z_{ni}A_{ni}
\left[ \frac{1}{2}+z_{ni}^{2}
\Big(1+i\sqrt{\pi}z_{ni} W\left( z_{ni}\right)   \Big)\right] 
\nonumber
\\ 
& \displaystyle 
+\sum\limits_{n=-\infty}^{\infty}\chi_{i}\eta_{i}
 \Big(1+i\sqrt{\pi}z_{ni} W\left( z_{ni}\right)   \Big) k^{2}_{\perp}\rho_{i}^{2}\left(A_{ni}+\hat{A}_{ni}\right)  
\nonumber
\\ & \displaystyle
\left.
+\tau\frac{v_{Te}}{v_{Ti}}\sum\limits_{n=-\infty}^{\infty}A_{ne}\left(z_{0e}-\chi_{e} \right)
\Big(1+i\sqrt{\pi}z_{ne}W\left( z_{ne}\right)   \Big)
\right\}=0
\nonumber 
\end{eqnarray}
\begin{figure}[!htbp]
\includegraphics[width=0.4\textwidth]{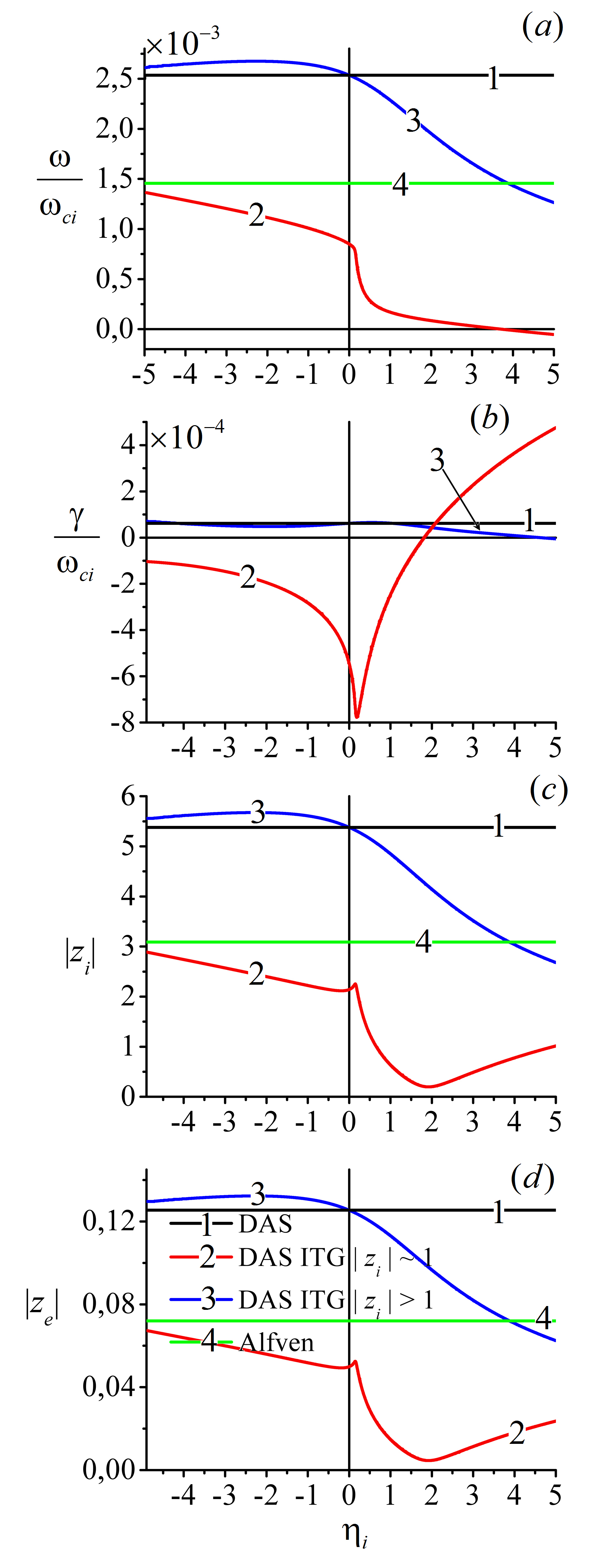}
\caption{\label{fig3} The normalized frequency $\omega/\omega_{ci}$ (panel (a)), the normalized growth rate $
\gamma/\omega_{ci}$ (panel (b)), $\left|z_{i} \right|$ (panel (c)) and $\left|z_{e} \right|$ (panel (d))
versus $\eta_{i}$  for $L_{n}/\rho_{i} = 35$, $k_{y}\rho_{i} = 0.1$, $T_{i}/T_{e} = 
1$, $\left( V'_{0}/\omega_{ci}\right) ^{-1} =70$, $\left( k_{z}\rho_{i}\right) ^{-1} = 3000$.}
\end{figure}
and
\begin{eqnarray}
& \displaystyle
\Phi\left(\mathbf{k},\omega \right) \left[
\sum\limits_{n=-\infty}^{\infty}A_{ni}\left(z_{0i}-\chi_{i}\left(1-\frac{1}{2} \eta_{i}\right) \right)
\right. 
\nonumber
\\ 
& \displaystyle 
\times
\Big(1+i\sqrt{\pi}z_{ni} W\left( z_{ni}\right)   \Big)
\nonumber
\\ 
& \displaystyle
-S_{i}\sum\limits_{n=-\infty}^{\infty}A_{ni}
z_{ni}\Big(1+i\sqrt{\pi}z_{ni} W\left( z_{ni}\right)   \Big)
\nonumber
\\ 
& \displaystyle 
-\sum\limits_{n=-\infty}^{\infty}\chi_{i}\eta_{i}z_{ni}A_{ni}
\left[ \frac{1}{2}+z_{ni}^{2}
\Big(1+i\sqrt{\pi}z_{ni} W\left( z_{ni}\right)   \Big)\right] 
\nonumber
\\ 
& \displaystyle 
+\sum\limits_{n=-\infty}^{\infty}\chi_{i}\eta_{i}
\Big(1+i\sqrt{\pi}z_{ni} W\left( z_{ni}\right)   \Big) k^{2}_{\perp}\rho_{i}^{2}\left(A_{ni}+\hat{A}_{ni}\right)  
\nonumber
\\ & \displaystyle
\nonumber
\\ & \displaystyle
\left.
+\tau\frac{v_{Te}}{v_{Ti}}\sum\limits_{n=-\infty}^{\infty}A_{ne}\left(z_{0e}-\chi_{e} \right)
\Big(1+i\sqrt{\pi}z_{ne}W\left( z_{ne}\right)   \Big)
\right]
\nonumber
\\ 
& \displaystyle
+A_{\parallel}\left(\mathbf{k},\omega \right)\left\{
\frac{k^{2}\rho_{i}}{2\beta}-\sum\limits_{n=-\infty}^{\infty}
\left(z_{0i}-\chi_{i}\left(1-\frac{1}{2} \eta_{i}\right) \right)
\right. 
\nonumber
\\ 
& \displaystyle
\times A_{ni}z_{ni}\Big(1+i\sqrt{\pi}z_{ni} W\left( z_{ni}\right)   \Big)
\label{6}
\\ & \displaystyle
+S_{i}\sum\limits_{n=-\infty}^{\infty}A_{ni}\left[\frac{1}{2} +z_{ni}^{2}
\Big(1+i\sqrt{\pi}z_{ni} W\left( z_{ni}\right)   \Big)\right] 
\nonumber
\\ 
& \displaystyle 
+\sum\limits_{n=-\infty}^{\infty}\chi_{i}\eta_{i}z_{ni}A_{ni}
\left[ \frac{1}{2}+z_{ni}^{2}
\Big(1+i\sqrt{\pi}z_{ni} W\left( z_{ni}\right)   \Big)\right] 
\nonumber
\\ 
& \displaystyle 
-\sum\limits_{n=-\infty}^{\infty}\chi_{i}\eta_{i}z_{ni} \:
 k^{2}_{\perp}\rho_{i}^{2}\left(A_{ni}+\hat{A}_{ni}\right)  
\nonumber
\\ 
& \displaystyle
\times 
 \Big(1+i\sqrt{\pi}z_{ni} W\left( z_{ni}\right)   \Big)
\nonumber
\\ 
& \displaystyle
-\tau\frac{v_{Te}^{2}}{v_{Ti}^{2}}\sum\limits_{n=-\infty}^{\infty}A_{ne}\left(z_{0e}-\chi_{e} \right)
\nonumber
\\ 
& \displaystyle 
\left.
\times
z_{ne}
\Big(1+i\sqrt{\pi}z_{ne}W\left( z_{ne}\right)   \Big)
\right\}=0.
\nonumber
\end{eqnarray}
\begin{figure}[!htbp]
\includegraphics[width=0.4\textwidth]{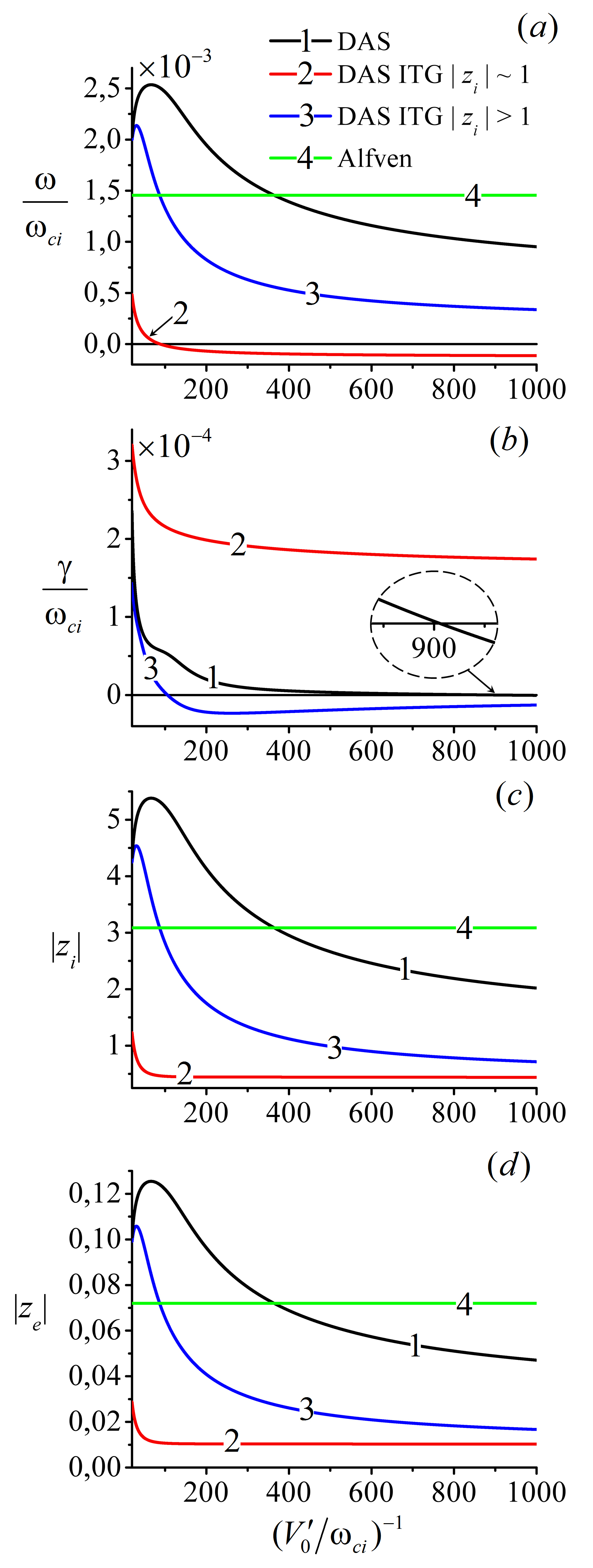}
\caption{\label{fig4} The normalized frequency $\omega/\omega_{ci}$ (panel (a)), the normalized growth rate $
\gamma/\omega_{ci}$ (panel (b)), $\left|z_{i} \right|$ (panel (c)) and $\left|z_{e} \right|$ (panel (d))
versus $\left( V'_{0}/\omega_{ci}\right) ^{-1}$  for $L_{n}/\rho_{i} = 35$, $\eta_{i}= 3$, $k_{y}\rho_{i} = 0.1$, $T_{i}/T_{e} = 
1$, $\left( k_{z}\rho_{i}\right) ^{-1} = 3000$.}
\end{figure}

In Eqs. (\ref{5}), (\ref{6})  $\lambda_{Di}$  is the ion Debye length, $\omega_{ci}$ is the ion cyclotron frequency,
$\rho_{i}= v_{Ti}/\omega_{ci}$  is the ion thermal Larmor radius, $A_{ni,e}=I_{n}\left(k_{\bot}^{2} \rho_{i,e}^{2} \right)e^{-
k_{\bot}^{2}\rho_{i,e}^{2}}$, $\hat{A}_{ni}=e^{-k_{\bot}^{2}\rho_{i}^{2}}I'_{n}\left(k_{\bot}^{2} \rho_{i}^{2} \right)$, $I_{n}$ is 
the modified Bessel function of order $n$, $z_{ni,e} = \left( \omega-n\omega_{ci,e}\right) /\sqrt{2}k_{z}v_{Ti,e}$,
$z_{0i,e} = \omega/\sqrt{2}k_{z}v_{Ti,e}=z_{i,e}$, $W\left( z \right)=e^{- z^{2}}\left(1
+\frac{2i}{\sqrt {\pi }}\int\limits_{0}^{z} e^{t^{2}}dt\right)$ is the complex error function, $\tau=T_{i}/T_{e}$, 
$S_{i}=k_{y}V'_{0}/k_{z}\omega_{ci}$, $v_{di,e}= \left( cT_{i,e}/eB_{0}\right)  \left( d\ln 
n_{i}/dx\right) $, $v_{di,e}$ is ion, electron diamagnetic velocity,    
$\eta_{i}= d\ln T_{i}/d\ln n_{i}$, $\chi_{i,e} = k_{y}v_{di,e}/\sqrt{2}k_{z}v_{Ti,e}$.
The principal difference of Eq. (\ref{6}) from the similar equation obtained 
early (see, for example Ref.\cite{Wang} and the references therein) consists in the accounting for the perturbed 
ion current, as well as the electron current.
We do not assume here, that the phase velocity of the perturbations 
is much above the ion thermal velocity. In the case $T_{i}\sim T_{e}$ and $\left|z_{0i}\right|\sim 1$ the electron and ion 
responses in Eq. (\ref{6}) are of the same order. 

We consider low frequency electromagnetic modes with frequency $\omega$ much less than the ion cyclotron 
frequency $\omega_{ci}$ in the limit $|\omega|\lesssim k_{z}v_{Te}$ as is appropriate for the
velocity shear and the temperature gradient instabilities. For these conditions, the general 
dispersion equation that accounts for the parallel--flow shear and inhomogeneous profiles of 
ion density and ion temperature and accounts for the effects of thermal motion of ions, both along and across the 
magnetic field, has a form 
\begin{eqnarray}
& \displaystyle 
A+ i\sqrt{\pi} W\left(z_{0i} \right) B =0, \label{7}
\end{eqnarray}
where
\begin{eqnarray}
& \displaystyle 
A = \left(1+iz_{0e}\sqrt{\pi}W\left(z_{0e} \right) \right)\left(\tau z_{0i}+\chi_{i} \right)
\nonumber
\\  & \displaystyle 
\times
\Big[\left(z_{0i}-\chi_{i} \right) \left(2A_{0i}-1\right)-z_{0i}A_{0i}\left( S_{i} + \eta_{i}\chi_{i}z_{0i}\right) 
\nonumber
\\ 	
& \displaystyle 
+2 \eta_{i}\chi_{i} \:k_{\bot}^{2} \rho_{i}^{2}\left( A_{0i}-A_{1i}\right) 
  \Big] 
\nonumber
\\ 	
& \displaystyle 
+\Big[1+\tau\left( 1+i\left(z_{0e}-\chi_{e} \right)\sqrt{\pi}
W\left(z_{0e} \right) \right)
\nonumber
\\ 	
& \displaystyle 
-A_{0i}\left( S_{i} + \eta_{i}\chi_{i}z_{0i}\right)  \Big]
\nonumber
\\ 
& \displaystyle \times
\left( \frac{k^{2}\rho^{2}_{i}}{2\beta}+\frac{S_{i}A_{0i}}{2}- A_{0i}z_{0i}
\left(z_{0i}Q_{i}-\chi_{i} \right)\right. 
\nonumber
\\ 
& \displaystyle	
\left.  + \eta_{i}\chi_{i}z_{0i} \Big[A_{0i}z_{0i}^{2}-k_{\bot}^{2} \rho_{i}^{2}\left( A_{0i}-A_{1i}\right) \Big] 
\right) 
\nonumber
\\ 
& \displaystyle	
+\Big[ A_{0i}\left(z_{0i}Q_{i}-\chi_{i} \right)\Big.
\nonumber
\\ 
& \displaystyle
\left. 
-\eta_{i}\chi_{i}z_{0i} \Big[A_{0i}z_{0i}^{2}-k_{\bot}^{2} \rho_{i}^{2}\left( A_{0i}-A_{1i}\right) \Big] \right] ^{2}
\label{8}
\end{eqnarray}
\begin{figure}[!htbp]
\includegraphics[width=0.4\textwidth]{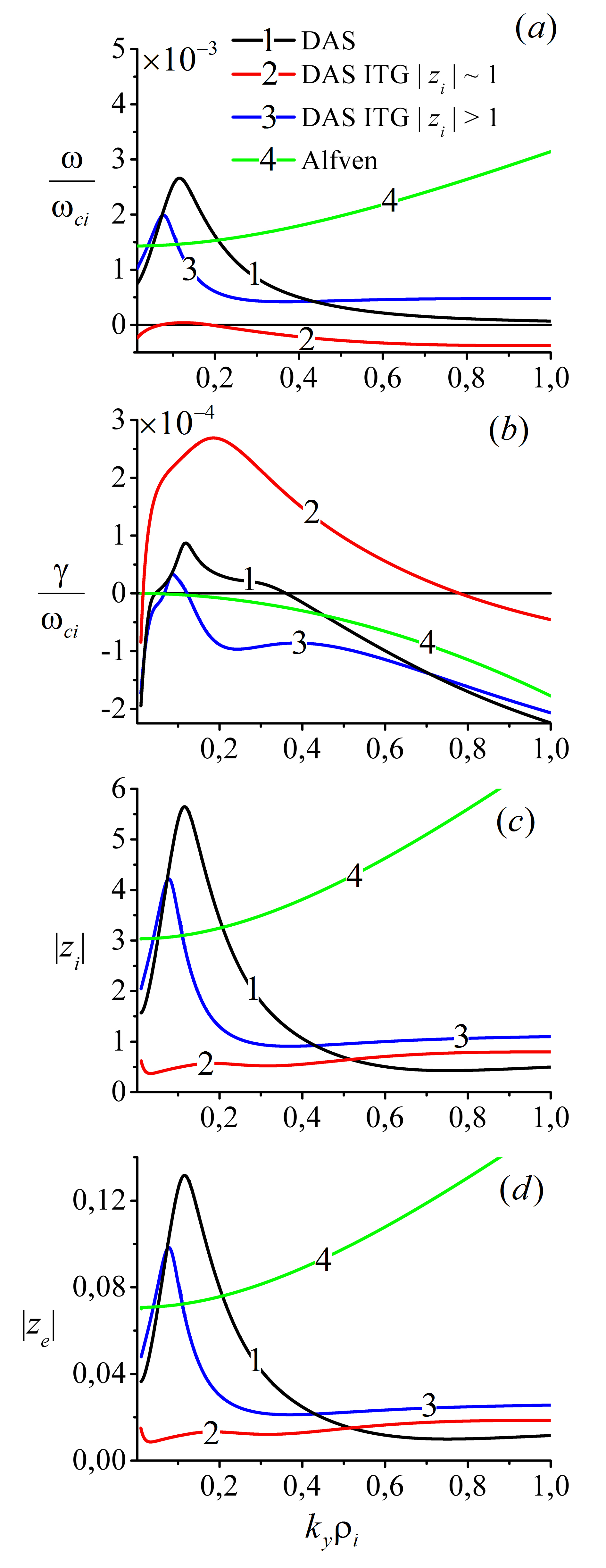}
\caption{\label{fig5} The normalized frequency $\omega/\omega_{ci}$ (panel (a)), normalized growth rate $
\gamma/\omega_{ci}$ (panel (b)), $\left|z_{i} \right|$ (panel (c)) and $\left|z_{e} \right|$ (panel (d))
versus  $k_{y}\rho_{i}$ for $L_{n}/\rho_{i} = 35$, $\eta_{i}= 3$, $\left( V'_{0}/\omega_{ci}\right) ^{-1} = 70$, $T_{i}/T_{e} = 
1$, $\left( k_{z}\rho_{i}\right) ^{-1} = 3000$.}
\end{figure}
and
\begin{eqnarray}
& \displaystyle 
B = \left( A_{0i} \left(z_{0i}Q_{i}-\chi_{i} \right) 
 +\eta_{i}\chi_{i}\left[A_{0i}\left(\frac{1}{2}-z_{0i}^{2} \right) \right. \right. 
 \nonumber
 \\ 
 & \displaystyle \left. \Big. +k_{\bot}^{2} \rho_{i}^{2}\left( A_{0i}-A_{1i}\right)\Big] \right) 
\nonumber
\\ 
& \displaystyle
\times\left[\frac{k^{2}\rho^{2}_{i}}{2\beta}+\frac{S_{i}A_{0i}}{2}-z_{0i}
\left(z_{0i}-\chi_{i} \right)\left(1-A_{0i} \right)\right. 
\nonumber
\\ 
& \displaystyle
\Big. 
+ \eta_{i}\chi_{i}z_{0i} \:k_{\bot}^{2} \rho_{i}^{2}\left( A_{0i}-A_{1i}\right) 	\Big]=0,
\label{9}
\end{eqnarray}
where $A_{0,1i}=I_{0,1}\left(k_{\bot}^{2} \rho_{i}^{2} \right)e^{-k_{\bot}^{2}\rho_{i}^{2}}$ and $Q_{i}=1-S_{i}$. In the case of the homogeneous ion 
temperature, i.e. for $\eta_{i}= 0$, Eq. (\ref{7}) becomes identical to dispersion equation obtained and analysed in Ref. \cite{Mikhailenko-2016}. In this 
paper, the numerical analysis of Eq. (\ref{7}) is performed with particular intent to ascertain the role of the ion temperature inhomogeneity on the DA 
instabilities in the presence of parallel sheared flow.

\section{Numerical solutions and discussions}\label{sec3}

\begin{figure}[!htbp]
\includegraphics[width=0.4\textwidth]{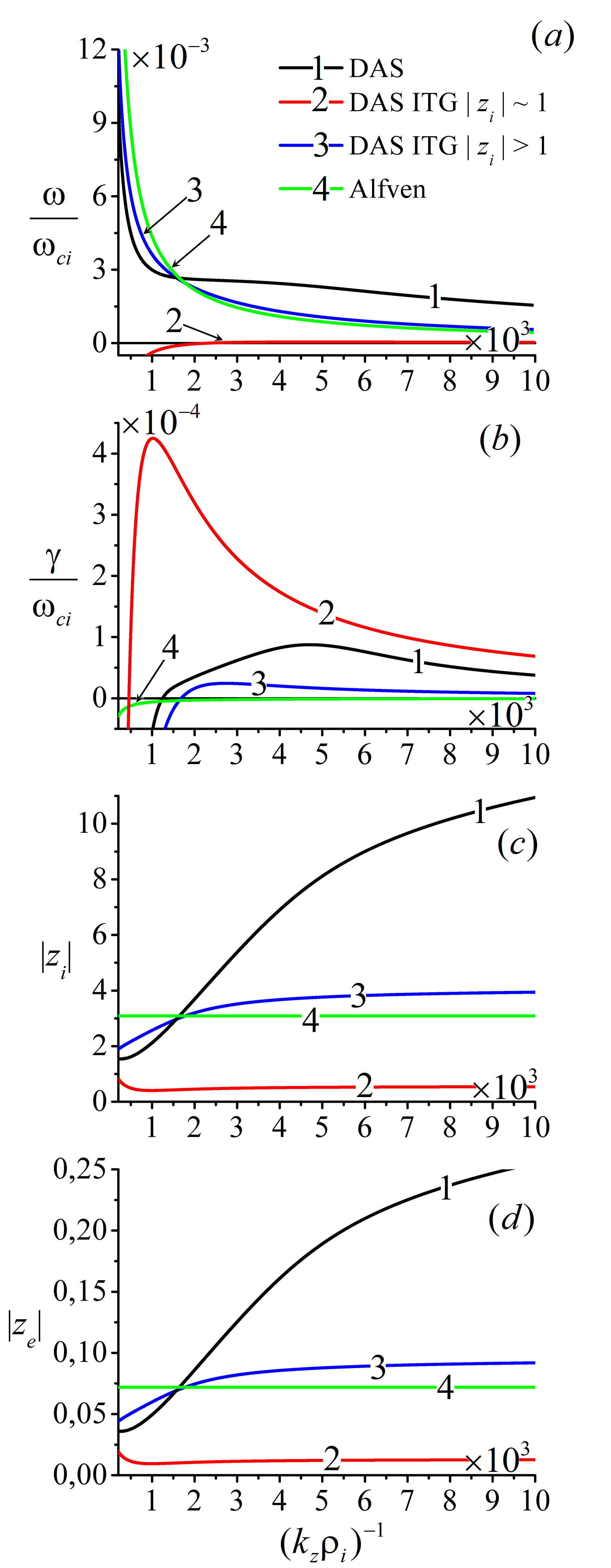}
\caption{\label{fig6} The normalized frequency $\omega/\omega_{ci}$ (panel (a)), the normalized growth rate $
\gamma/\omega_{ci}$ (panel (b)), $\left|z_{i} \right|$ (panel (c)) and $\left|z_{e} \right|$ (panel (d))
versus $\left( k_{z}\rho_{i}\right) ^{-1}$  for $L_{n}/\rho_{i} = 35$, $\eta_{i}= 3$, $k_{y}\rho_{i} = 0.1$, $\left( V'_{0}/\omega_{ci}\right) ^{-1} 
= 70$, $T_{i}/T_{e} = 1$ and $\beta = 5\%$.}
\end{figure}

\begin{figure}[!htbp]
\includegraphics[width=0.4\textwidth]{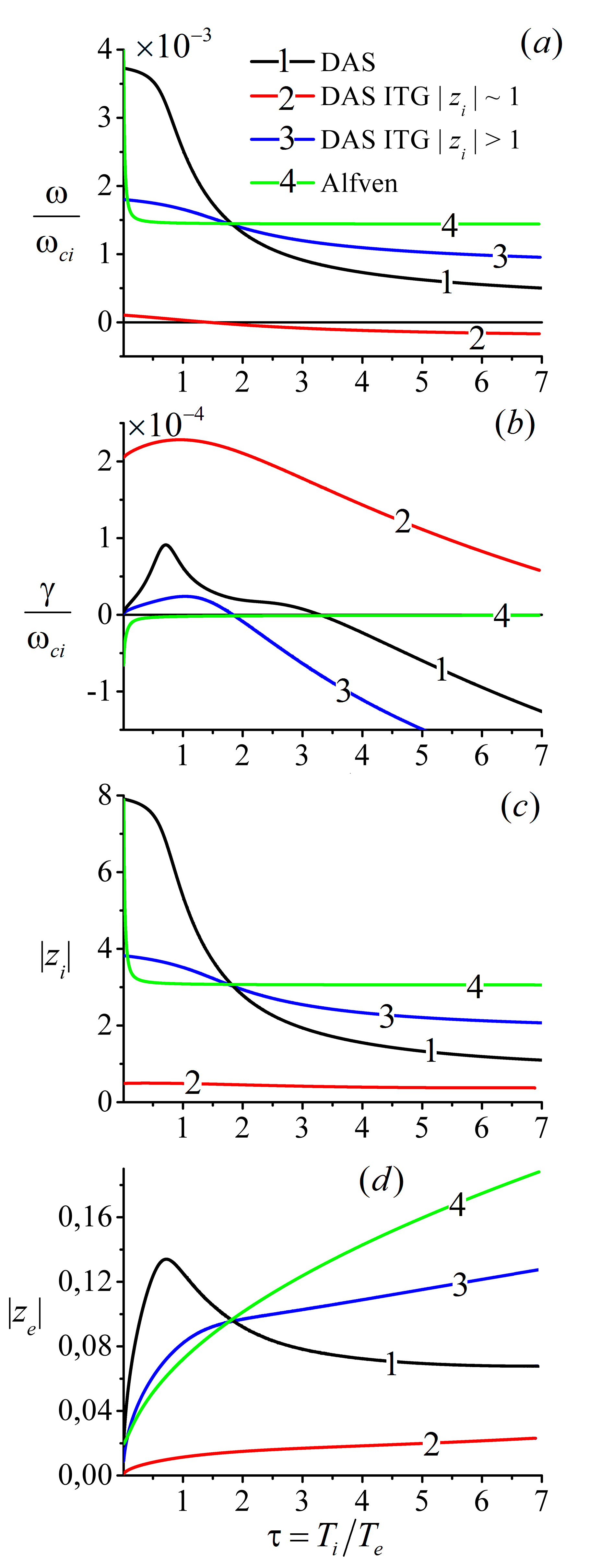}
\caption{\label{fig7} The normalized frequency $\omega/\omega_{ci}$ (panel (a)), the normalized growth rate $
\gamma/\omega_{ci}$ (panel (b)), $\left|z_{i} \right|$ (panel (c)) and $\left|z_{e} \right|$ (panel (d))
versus $\tau = T_{i}/T_{e}$  for $L_{n}/\rho_{i} = 35$, $\eta_{i}= 3$, $k_{y}\rho_{i} = 0.1$, $\left( V'_{0}/\omega_{ci}\right) ^{-1} = 70$,
$\left( k_{z}\rho_{i}\right) ^{-1} = 3000$ and $\beta = 5\%$.}
\end{figure}

In a general case, all terms in Eq. (\ref{7}) are of the same order of value and the numerical analysis of Eq. (\ref{7}) is necessary. 
The results of the numerical solution of Eq. (\ref{7}) are presented in Figs. \ref{fig1}-\ref{fig7}. In all these Figures (with the exception of Fig. 
\ref{fig2}) the results of the numerical solution  for the normalized 
frequency $\omega/\omega_{ci}$ are presented in panel (a), for the the normalized growth rate $\gamma/\omega_{ci}$ in panel (b), 
for the parameter $\left|z_{i} \right|$ in panel (c), and for $\left|z_{e}\right|$ in panel (d). These solutions were derived for 
a plasma with $\beta = 5\%$ and $m_{i}/m_{e}=1840$.

In Fig. \ref{fig1}, the solution to equation (\ref{7}) is given versus $L_{n}/\rho_{i}$ for  $\eta_{i} = 3$, $k_{y}\rho_{i} = 0.1$, 
$T_{i}/T_{e} = 1$, $\left( V'_{0}/\omega_{ci}\right) ^{-1}=70$, $\left( k_{z}\rho_{i}\right)^{-1} = 3000$. In this figure, 
as well as in all others, the plots for the  kinetic Alfven wave in a steady 
plasma are denoted by a green line (line 4), the plots for the DA instability in the shear flow (DAS instability) with homogeneous 
ion temperature ( i. e. for $\eta_{i}=0$) are denoted by a black line (line 1). It was found in Ref.\cite{Mikhailenko-2016} that the solution to Eq. 
(\ref{7}) for $\eta_{i}=0$ reveals that the velocity shear is the factor which modifies the dispersion properties of the DA instability 
which develops\cite{Mikhailovskii} in a steady plasma with cold ions ($T_{i}\ll T_{e}$) and is the source of the free energy for the development 
of other DAS instability which is absent in a steady plasma. It was derived \cite{Mikhailenko-2016} that the left part of line 1 in Fig. 
\ref{fig1} (panel (b)), which 
involves first localised maximum of the growth rate, corresponds to the shear-flow modified DAS instability. This instability develops as the DA 
instability due to the inverse electron Landau damping, however exists in the plasma with 
comparable inhomogeneous ion temperature. The right part of the line 1 with $L_{n}/\rho_{i}>75$ corresponds to the 
shear flow driven DAS instability which develops due to the coupled action of the ion Landau damping and flow velocity shear. It was named in Ref. 
\cite{Mikhailenko-2016} as the ion kinetic shear flow driven DAS instability. This instability  has the growth rate larger than the growth rate of the 
modified DAS instability. It exists in the parallel sheared flow and is absent in the uniform plasma flow or steady plasmas. As it is presented in 
Figs. \ref{fig1}--\ref{fig7}, these two DAS instabilities develop at different plasma parameters and at the different ranges of the wave 
number values.

The numerical solution to Eq. (\ref{7}) with $\eta_{i}\neq 0$ predicts the development of two distinct DA instabilities in parallel sheared flow with 
inhomogeneous ion temperature (DAS-ITG instabilities). The plots for the frequency and the growth rate of the first DAS-ITG instability are 
denoted  by a red line (line 2) and the plots for the second DAS-ITG instability are denoted by a blue line (line 3) in all 
Figs. \ref{fig1} -- \ref{fig7}. The panels (a), (b), (c) (d) of Figs. \ref{fig1}, \ref{fig3}--\ref{fig7} display that this instability 
develops under the conditions of the strong inverse ion Landau damping ($|z_{i}|\lesssim 1$) with the growth rate 
$\gamma\left(\mathbf{k}\right)$ of the order of the frequency $\omega\left(\mathbf{k}\right)$ and the useful estimate for the growth rate,
\begin{eqnarray}
& \displaystyle 
\gamma \sim k_{z}v_{Ti}, \label{10}
\end{eqnarray}
follows. This instability may be named as the ion kinetic DAS-ITG instability. 

The second DAS-ITG instability exists in the finite domain of the parameter $L_{n}/\rho_{i}$ where the frequency of this instability and of the kinetic 
Alfven wave are almost equal (Fig. \ref{fig2} (panel b)) and where the ion Landau damping is weak (i. e. where $|z_{i}|\gg 1$ (Fig. \ref{fig1} (panel (c))). 
This instability develops due to the inverse electron Landau damping with the growth rate (Fig. \ref{fig1} (panel b)) much less than the frequency (panel 
(a)). It is the modified version of the DA instability of the steady plasmas\cite{Mikhailovskii}. In the parallel sheared flow, both DAS-ITG instabilities 
develop, as it follows from Fig. \ref{fig7}, in the plasma with cold as well as with hot ions where $T_{i}
\gtrsim T_{e}$. The second DAS-ITG instability exists, however, only when the flow velocity shear is sufficiently strong (see Fig. \ref{fig4} (panel (b))). 
Therefore this instability may be named as the electron kinetic shear flow modified DAS-ITG instability.  As it follows from panel (b) of Fig. \ref{fig2}, 
the frequency of this instability in the parameters regions where the growth rate is maximum may be estimated as
\begin{eqnarray}
& \displaystyle 
\omega \sim k_{y}v_{de}. \label{11}
\end{eqnarray}
Figure \ref{fig1} displays that for the used values of the parameters the ion kinetic DAS-ITG instability develops for any values of $L_{n}/\rho_{i}$, 
whereas the electron kinetic shear flow modified DAS-ITG instability exists in the finite interval $20 \lesssim L_{n}/\rho_{i} \lesssim 80$. In Figs.  
\ref{fig3} --\ref{fig7}, we use the value  $L_{n}/\rho_{i} =35$ for which both instabilities exists and both may be investigated.

In Fig. \ref{fig3}, the solution to equation (\ref{7}) is given versus $\eta_{i}$  for $L_{n}/\rho_{i} = 35$, $k_{y}\rho_{i} = 0.1$, $T_{i}/T_{e} = 1$, 
$\left(V'_{0}/\omega_{ci}\right)^{-1} =70$, $\left(k_{z}\rho_{i}\right) ^{-1} = 3000$. This figure displays that the ion temperature 
inhomogeneity affects  differently on these DAS-ITG instabilities. The growth rate of the electron kinetic sheared flow modified DAS-ITG instability 
gradually decay with  parameter $\eta_{i}\sim L_{n}/L_{Ti}$ growth. At the same time, the ion temperature inhomogeneity has a decisive effect on the 
development of the ion kinetic DAS-ITG instability. This instability develops when parameter $\eta_{i}$ becomes larger the threshold value with the growth 
rate growing with $\eta_{i}$ value growth. Therefore, in fact only the ion kinetic DAS-ITG instability may be considered as a ion-temperature-gradient-driven 
instability. For $\eta_{i}=3$ value, which is used in Figs. \ref{fig1}, \ref{fig4}--\ref{fig7}, both DAS-ITG instabilities exist.

In Fig. \ref{fig4}, the solution to equation (\ref{7}) is given 
versus $\left( V'_{0}/\omega_{ci}\right) ^{-1}$  for $L_{n}/\rho_{i} = 35$, $\eta_{i}= 3$, $k_{y}\rho_{i} = 0.1$, $T_{i}/T_{e} = 
1$, $\left( k_{z}\rho_{i}\right) ^{-1} = 3000$.  Figure \ref{fig4} displays, that the growth rates of both instabilities grow with growth 
of the flow velocity shearing rate. In the plasma with equal ion and electron temperatures, the electron kinetic shear flow modified 
DAS-ITG instability exists only in the limited range of the sufficiently 
large velocity shear. For the used numerical values of other parameters this instability exists only when $V'_{0}> 10^{-2}\omega_{ci}$.
The ion kinetic DAS-ITG instability continues to exist at any velocity shear values and stems from the coupled action of the flow 
velocity shear, ion temperature gradient and inverse ion Landau damping ($\left|z_{i}\right|\approx 1.2\div 0.5$, $\left|z_{e}\right|\approx 10^{-2}$
(panels (c) and (d)). For $\left( V'_{0}/\omega_{ci}\right) ^{-1} = 70$ used in Figs. \ref{fig1}--\ref{fig7} (with the exception of Fig. 
\ref{fig4}) both DAS-ITG instabilities develop.

In Fig. \ref{fig5}, the solution to Eq. (\ref{7}) is given versus $k_{y}\rho_{i}$ for $L_{n}/\rho_{i}= 35$, $\eta_{i} = 3$, 
$\left( V'_{0}/\omega_{ci}\right) ^{-1} = 70$,
$T_{i}=T_{e}$, $\left(k_{z}\rho_{i} \right)^{-1}= 3000$. This figure reveals, that for
the employed parameters the electron kinetic DAS-ITG instability develops in the narrow
interval $\backsim 0.05 \div 0.15$ of the $k_{y}\rho_{i}$ values with the growth rate much less than the frequency.
At the same parameters, the ion kinetic DAS-ITG instability develops with the growth
rate much above the frequency in the wide interval $\backsim 0.05 \div 0.8$ of the $k_{y}\rho_{i}$ values. Note,
that the frequency of the ion kinetic DAS-ITG instability changes its sign in this interval:
it is positive in the narrow region of $k_{y}\rho_{i}$ where the growth rate is maximum being negative at the rest part 
of the  $k_{y}\rho_{i}$ values. It follows from Fig. \ref{fig5} that both DAS-ITG instabilities exist for the $k_{y}\rho_{i}= 0.1$ 
value used in the calculations presented in all other Figures.

In Fig. \ref{fig6}, the solution to Eq. (\ref{7}) is given versus $\left( k_{z}\rho_{i}\right) ^{-1}$  
for $L_{n}/\rho_{i} = 35$, $\eta_{i}= 3$, $k_{y}\rho_{i} = 0.1$, $\left( V'_{0}/\omega_{ci}\right) ^{-1}=70$, 
$T_{i}/T_{e} = 1$.  The panel (b) displays 
that the ion kinetic DAS-ITG instability has largest growth rate in comparison with
electron kinetic DAS-ITG instability and with DAS instability of a plasma with homogeneous
ion temperature for all considered values of $\left( k_{z}\rho_{i}\right) ^{-1}$. More over, the maximum growth
rate of the ion kinetic DAS-ITG instability attains for $\left( k_{z}\rho_{i}\right) ^{-1}$ values 
$\left( \left( k_{z}\rho_{i}\right) ^{-1}\lesssim 10^{-3}\right)$ for which the electron kinetic DAS-ITG instability as well as the considered in  
Ref.\cite{Mikhailenko-2016} DAS instabilities of the parallel sheared flow with homogeneous ion temperature are absent. In the calculations presented in all 
other figures we used $\left(k_{z}\rho_{i} \right)^{-1}= 3000$ value for which both DAS-ITG instabilities exist and the ion kinetic instability has the 
growth rate approximately in two time less than the maximum value.

In Fig. \ref{fig7}, the solution to Eq. (\ref{7}) is given versus $T_{i}/T_{e}$  for $L_{n}/\rho_{i} = 35$, 
$\eta_{i}= 3$, $k_{y}\rho_{i} = 0.1$, $\left( V'_{0}/\omega_{ci}\right) ^{-1}=70$, $\left( k_{z}\rho_{i}
\right) ^{-1}= 3000$.  This figure
displays that the DAS instability considered in Ref.\cite{Mikhailenko-2016}, and the electron and ion kinetic DAS-ITG instabilities develop
in parallel sheared flow with warm ions. It is contrary to the steady plasmas where the DA instability may be developed only in plasma 
with $T_{i}\ll T_{e}$. The maximum growth rate attains for the plasma
with $T_{i}\simeq T_{e}$ (DAS instability) or when $T_{i}\gtrsim T_{e}$ (both DAS-ITG instabilities). Panel
(b) demonstrates that DAS instabilities and electron kinetic DAS-ITG instability exist in the finite
ion/electron temperatures ratio interval which involves $T_{i}/T_{e}=1$ value, used in all other Figures. The ion kinetic DAS-ITG instability exists for all considered values of the temperatures ratio.

\section{Conclusions}\label{sec4}

The numerical analysis of the dispersion equation (\ref{7}), which accounts for the parallel flow shear, the inhomogeneous profiles of the 
plasma density and of the ion temperature, and the effects of thermal motion of ions was performed. It was derived, that the 
parallel sheared flow of a plasma with inhomogeneous ion temperature is unstable against the 
development of two distinct DAS-ITG instabilities. 
The performed numerical analysis of the dispersion equation (\ref{7}) displays the existence of the shear flow driven ion kinetic DAS-ITG instability which 
develops due to the combined action of the flow velocity shear, ion temperature gradient and ion Landau damping. This instability has the growth rate of the 
order of the frequency. This growth rate is also much above the frequency of the electron kinetic DAS-ITG instability, which has 
the growth rate much less than the frequency. 

Thus, the DA turbulence, which is powered by the DA instability existing only in a steady plasma with cold 
ions, $T_{i}\ll T_{e}$, relives in parallel sheared plasma flows with inhomogeneous ion temperature of the order of or above the electron 
temperature as DAS-ITG turbulence with much more larger growth rate. The initially fastest growing disturbance due to the ion kinetic DAS-ITG 
instability will dominate the subsequent development the DA turbulence in parallel shear flow. Certainly, because $\gamma\left(\mathbf{k}\right)\sim \omega
\left(\mathbf{k}\right)$ for this instability the nonlinear analysis of this 
instability with can't be performed on the base of the weak turbulence approach. Also, the renormalized nonlinear theory\cite {Mikhailenko-2016.1}, which 
accounts for the scattering of ions by the DAS turbulence, for which $\gamma\left(\mathbf{k}\right)\sim \omega
\left(\mathbf{k}\right)$ also\cite {Mikhailenko-2016.1}, may give analytically only the approximate 
estimate for the saturated amplitude of the electric field. The phase randomization of the waves in the wave packet with wave number spectrum 
width $\Delta k_{\bot}\sim k_{\bot} $ and the frequency $\omega \left( \mathbf{k}\right) $ occurs at time 
$t\sim  \gamma^{-1} \left( \mathbf{k}\right) $ when 
\begin{eqnarray}
&\displaystyle 
\frac{d \omega \left( \mathbf{k}\right)}{dk_{\bot}}\Delta k_{\bot}t \sim \frac{\omega \left( \mathbf{k}\right)}
{\gamma \left( \mathbf{k}\right)}\gtrsim \pi,
\label{12}
\end{eqnarray}
but this does not occur for the ion kinetic DAS-ITG instability. There is no small parameter which can be applied for the development of any kind of 
the turbulence theory. The simplest estimates for the saturated amplitude of the electric 
field of the DAS-ITG turbulence powered by the considered DAS-ITG instabilities may be derived at the most general level by employing the widely invoked  
"mixing length estimate".  It balances the wavelength of the perturbation against the particle (in our case the ion) displacement, $\xi$,  in the unstable 
electric field, 
\begin{eqnarray}
&\displaystyle 
\xi \sim \frac{u}{\gamma}\sim \frac{2\pi}{k_{\perp}}.
\label{13}
\end{eqnarray}
Eq. (\ref{13}) defines the threshold for stochastization or mixing of a test ion trajectory and is the familiar for the instability 
saturation level\cite{Diamond}. With the estimates (\ref{10}) for the growth rate $\gamma \sim k_{z}v_{Ti}$ and for the ion velocity $u \sim 
c E/B\sim ck_{\bot}\varphi/B$, Eq. (\ref{13}) gives the estimate
\begin{eqnarray}
&\displaystyle 
\frac{e\varphi}{T_{i}}\sim \frac{k_{z}}{k_{\bot}}\frac{2\pi}{k_{\bot}\rho_{i}}.
\label{14}
\end{eqnarray}
This estimate is the same as was obtained in Ref.\cite{Mikhailenko-2016.1} for the perturbed potential in the steady state of the ion kinetic shear flow 
driven DAS instability of a plasma with homogeneous ion temperature. Note, that the estimate (\ref{14}) was obtained in Ref.
\cite{Mikhailenko-2016.1} employing the renormalized nonlinear theory, which accounts for the scattering of ions by the 
ensemble of DAS waves with random phases. 

The mixing length estimate may be employed for the estimating the steady state level for the electron kinetic DAS-ITG instability. In this case the
balance equation has a form
\begin{eqnarray}
&\displaystyle 
\xi \sim \frac{u}{\omega}\sim \frac{2\pi}{k_{\perp}},
\label{15}
\end{eqnarray}
where the estimate for the frequency $\omega$ is given by Eq. (\ref{11}).  The estimate for the steady state level for the perturbed potential,
\begin{eqnarray}
&\displaystyle 
\frac{e\varphi}{T_{i}}\lesssim \frac{1}{k_{\bot}L_{n}},
\label{16}
\end{eqnarray}
appeared to be the same as the obtained in Ref.\cite{Mikhailenko-2016.1} for the electron kinetic shear flow modified  DAS instability.

\begin{acknowledgments}
This work was supported  by National R$\&$D Program through the National
Research Foundation of Korea(NRF) funded by the Ministry of Education, Science and 	Technology (Grant No. NRF-2014M1A7A1A03029878) and BK21 Plus Creative 
Human Resource Development Program for IT Convergence.
\end{acknowledgments}

\bigskip
{\bf DATA AVAILABILITY}

\bigskip
The data that support the findings of this study are available from the corresponding
author upon reasonable request.

\end{document}